\documentclass[a4paper,11pt]{article}
\usepackage{pos}
\usepackage{wrapfig}
\usepackage{enumitem}
\usepackage{xspace}

\newcommand{\apj}{ApJ}
\newcommand{\apjs}{ApJS}

\newcommand{\apjl}{ApJL}

\newcommand{\aap}{A\&A}

\newcommand{\jcap}{J. Cosmology Astropart. Phys.}

\newcommand{\mnras}{MNRAS}

\newcommand{\prd}{Phys.~Rev.~D}

\newcommand{\prl}{Phys.~Rev.~Letter}

\newcommand{\ssr}{S Review}

\title{Cosmographic model of the astroparticle skies}
\ShortTitle{Cosmographic model of the astroparticle skies}

\author*[a]{Jonathan Biteau}
\author[a]{Sullivan Marafico}
\author[a]{Younes Kerfis}
\author[a]{Olivier Deligny}

\affiliation[a]{Universit\'e Paris-Saclay, CNRS/IN2P3, IJCLab, 91405 Orsay, France}

\emailAdd{biteau@in2p3.fr}

\abstract{Modeling the extragalactic astroparticle skies involves reconstructing the 3D distribution of the most extreme sources in the Universe. Full-sky tomographic surveys at near-infrared wavelengths have already enabled the astroparticle community to bind the density of sources of astrophysical neutrinos and ultra-high cosmic rays (UHECRs), constrain the distribution of binary black-hole mergers and identify some of the components of the extragalactic gamma-ray background. This contribution summarizes the efforts of cleaning and complementing the catalogs developed by the gravitational-wave and near-infrared communities, in order to obtain a cosmographic view on stellar mass ($M_*$) and star formation rate (SFR). Unprecedented cosmography is offered by a sample of about 400,000 galaxies within 350 Mpc, with a 50-50 ratio of spectroscopic and photometric distances, $M_*$, SFR and corrections for incompleteness with increasing distance and decreasing Galactic latitude. The inferred 3D distribution of $M_*$ and SFR is consistent with Cosmic Flows. The $M_*$ and SFR densities converge towards values compatible with deep-field observations beyond 100 Mpc, suggesting a close-to-isotropic distribution of more distant sources. In addition to highlighting relevant applications for the four astroparticle communities, this contribution explores the distribution of $B$-fields at Mpc scales deduced from the 3D distribution of matter, which is believed to be crucial in shaping the ultra-high-energy sky. These efforts provide a new basis for modeling UHECR anisotropies, which bodes well for the identification of their long-sought sources.}

\FullConference{37$^{\rm{th}}$ International Cosmic Ray Conference (ICRC 2021)\\
		July 12th -- 23rd, 2021\\
		Online -- Berlin, Germany}

\begin{document}
\maketitle

\section{Astroparticles and full-sky catalogs}\vspace{-0.2cm}

Astroparticle physics has emerged as a full-fledged branch of astronomy since the beginning of the 21st century. Largely driven by the particle-physics community, tremendous improvements in the sensitivity and angular resolution of astroparticle observatories have enabled the detection by \textit{Fermi}-LAT of thousands of extragalactic $\gamma$-ray sources at GeV energies \cite{2020ApJ...892..105A} and of nearly a hundred ones at TeV energies by H.E.S.S., MAGIC and VERITAS.\footnote{\url{http://tevcat2.uchicago.edu/}} Flares of active galactic nuclei (AGN) have been claimed as possible sources of PeV neutrinos detected by IceCube \cite{2018Sci...361.1378I}. Ultra-high energy cosmic rays (UHECR) observed by the Pierre Auger Observatory at multi EeV energies appear to be distributed on the sphere consistently with baryonic matter within a hundred Mpc \cite{2018ApJ...868....4A}. The latest messenger of astroparticle physics, in the form of gravitational waves (GW), unveiled about fifty binary mergers detected by LIGO and Virgo up to the end of 2019 \cite{2021PhRvX..11b1053A}. 

The identification of extragalactic astroparticle sources heavily relies on multi-wavelength observations. As the combination of radio, optical and X-ray observations enabled,  during the second half of the 20th century, a physical understanding of the nature e.g.\ of AGNs, the cross-correlations of datasets established by conventional astronomy with those collected by astroparticle observatories is unfolding before us a multi-messenger era that promises a renewed view on the non-thermal universe. Deep-field observations, combined with spectroscopic and photometric redshift measurements, have for example unveiled GeV-emitting blazars and $\gamma$-ray bursts (GRB) at distances beyond the cosmic noon, corresponding to the peak of the cosmic star formation history at $z \approx 2-3$, i.e.\ at a lookback time of ${\sim}\,10\,$Gyr. Nonetheless, the full-sky picture that astroparticle physicists employed for population and cross-correlation studies (e.g.\ \cite{2020JCAP...07..042A} and \cite{2018ApJ...853L..29A} for neutrinos and UHECRs) mostly remained limited over the past decade to 140\,Mpc ($z \approx 0.03$ or ${\sim}\,400\,$Myr). Such a distance limit corresponds to the volume probed by the 2MASS redshift survey (2MRS, \cite{2012ApJS..199...26H}) through measurements of spectroscopic distances of galaxies and estimates of their stellar mass ($M_*$) at near-infrared (NIR) wavelengths. The 2MASS photometric redshift catalog (2MPZ, \cite{2014ApJS..210....9B}) now provides distance estimates with a precision of ${\sim}\,12\%$ down to seven times lower fluxes in a fifteen times larger volume out to 350\,Mpc, that is $z\approx 0.08$ or 1\,Gyr. Such a reach was exploited e.g.\ to constrain the contributions of AGNs, star-forming galaxies and dark matter to the extragalactic $\gamma$-ray background \cite{2018PhRvD..98j3007A}. The 2MPZ catalog was recently complemented by the GW community \cite{2018MNRAS.479.2374D, 2020MNRAS.492.4768D} to foster galaxy-targeting approaches, which enable swift follow-up observations of binary neutron-star mergers, and to pave the way to accurate statistical constraints on the Hubble constant, assuming that binary black-hole mergers are traced by the $M_*$ distribution \cite{2021ApJ...909..218A}.

We further expanded in Ref.~\cite{2021arXiv210511345B} the catalogs developed by the GW and NIR communities to account for incompleteness as a function of luminosity distance, limited by the 2MASS sensitivity threshold, and of angular distance to the Galactic plane, where source confusion and obscuration mask background galaxies in the so called Zone of Avoidance (ZoA). Our publicly available catalog\footnote{\url{http://doi.org/10.5281/zenodo.4783406}} offers an unprecedented view on $M_*$ and star formation rate (SFR) on scales ranging from the very outskirts of the Milky Way out to the most distant superclusters identified in Cosmic Flows \cite{2017NatAs...1E..36H}. Of interest to the four astroparticle messengers, the use of this catalog is exemplified hereafter by a modeling of the UHECR sky.

\section{A cosmographic view on stellar mass and star formation}\vspace{-0.2cm}

\begin{figure}[b]
    \centering
    \includegraphics[width=.58\textwidth]{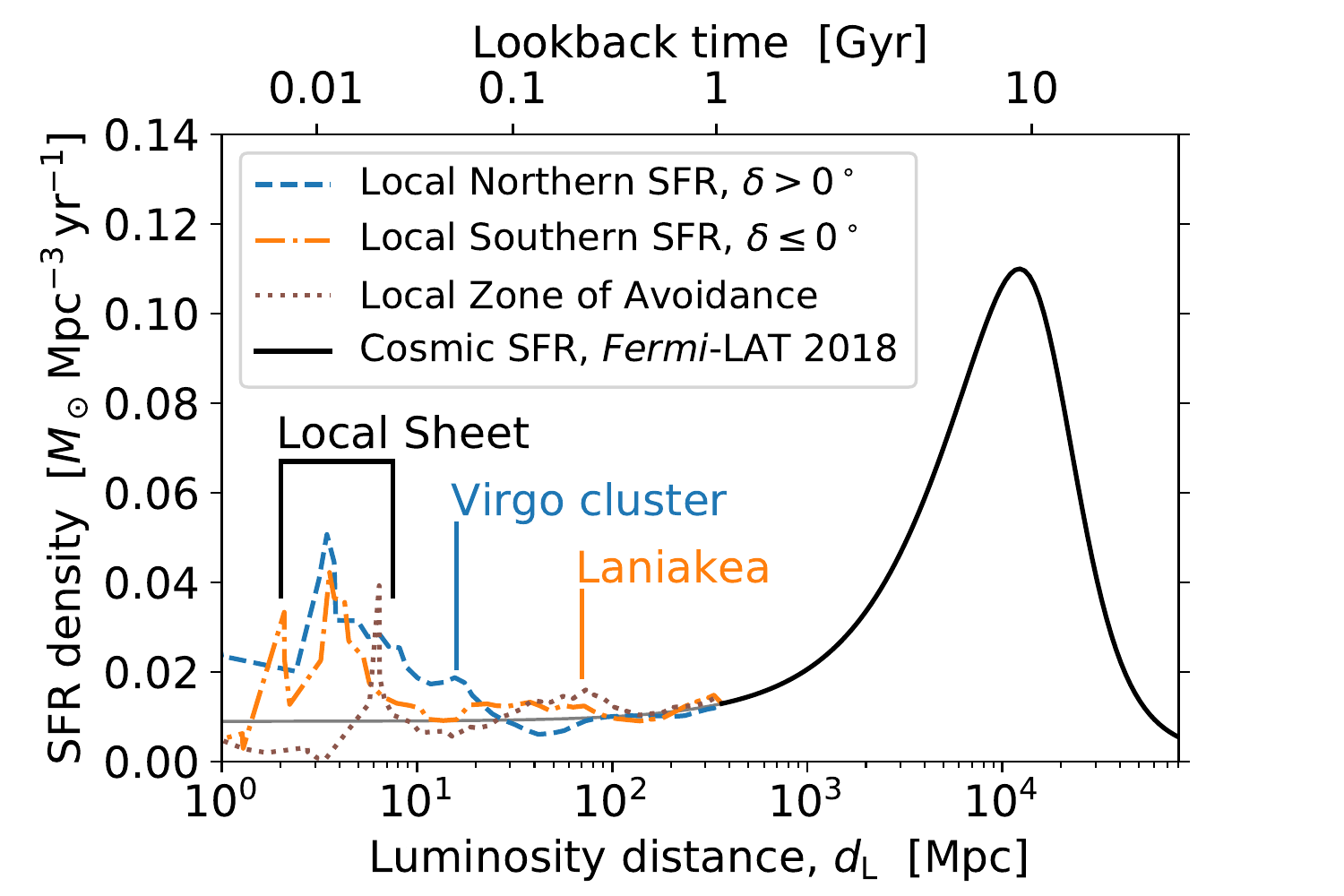} 
    \caption{The star formation history from cosmic dawn down to the past billion year. The SFR density at lookback times larger than 1\,Gyr, shown as a black solid line, follows the \textit{Fermi}-LAT constraints parametrized by Ref.~\cite{2021MNRAS.503.2033K} (see Table~5, initial SFR normalization). The local overdensity at shorter lookback times in the Northern and Southern hermispheres as well as in the Zone of Avoidance are estimated from our catalog and illustrated by colored lines, as labeled in the figure.\vspace{-0.5cm}}
    \label{fig:sfrd}
\end{figure}

The catalog established in Ref.~\cite{2021arXiv210511345B}  comprises 410,761 galaxies out to 350\,Mpc, distance at which 50\% of $M_*$ is below the 2MASS sensitivity limit. We have checked the distance of each galaxy against those tabulated in the HyperLEDA database \cite{2014A&A...570A..13M}, which provides cosmic-ladder estimates for about 4,000 nearby galaxies that are not in the Hubble flow, spectroscopic estimates for about half of the sample and 2MPZ photometric estimates for the other half. The crossmatch with HyperLEDA improves the number of spectroscopic estimates by a factor of four with respect to catalogs established by the GW community \cite{2018MNRAS.479.2374D, 2020MNRAS.492.4768D}. The sample is compatible with being flux limited at high Galactic latitudes but a deficit of galaxies is observed in the ZoA. The galaxy-count decrease in the ZoA is modeled with an empirical function of Galactic latitude, which provides incompleteness correction factors. Such corrections do not suffice close to the Galactic bulge and we employ galaxy cloning, i.e.\ the filling of the ZoA with galaxies from mirrored regions above and below the Galactic plane, to establish a realistic model for the entire sky. The $M_*$ distribution derived from the $4\pi$ sample is shown to be consistent with the constraints derived from deep-field observations.  The parametrization of the deep-field $M_*$ function is used to infer incompleteness correction factors as a function of luminosity distance. Accounting for such corrections on the sphere and in depth, the $M_*$ density is shown to converge towards the deep-field value for distances larger than 100\,Mpc. SFR estimates are also provided for each galaxy in the sample, exploiting the relation between $M_*$ and SFR for three morphological branches established with NIR and H$\alpha$ observations of galaxies in the Local Volume, at distances smaller than 11\,Mpc. Morphological information is available from HyperLEDA for about a third of the sample out to 350\,Mpc and the observed morphological distribution as a function of distance is exploited to provide a statistical estimate of SFR for the remaining two thirds of the sample. The catalog finally provides incompleteness correction factors for the SFR density, which is shown, as for $M_*$, to converge beyond 100\,Mpc towards the deep-field value. The SFR densities in the Northern and Southern hemispheres, as well as in the ZoA, connect smoothly with that at lookback times larger than 1\,Gyr (see Fig.~\ref{fig:sfrd}).

\begin{figure}[t]
    \centering
    \includegraphics[width=.45\textwidth]{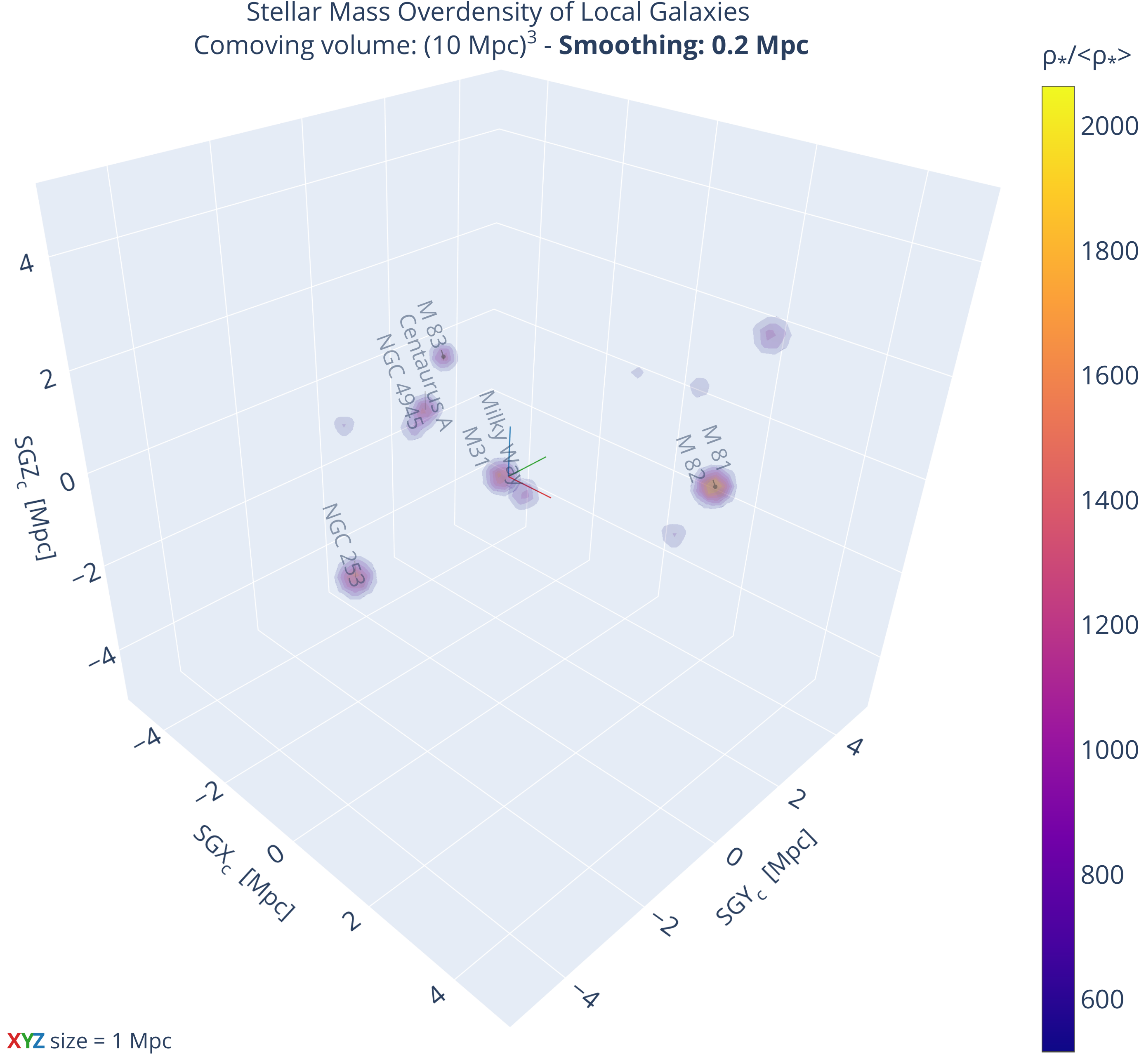} \hfill \includegraphics[width=.45\textwidth]{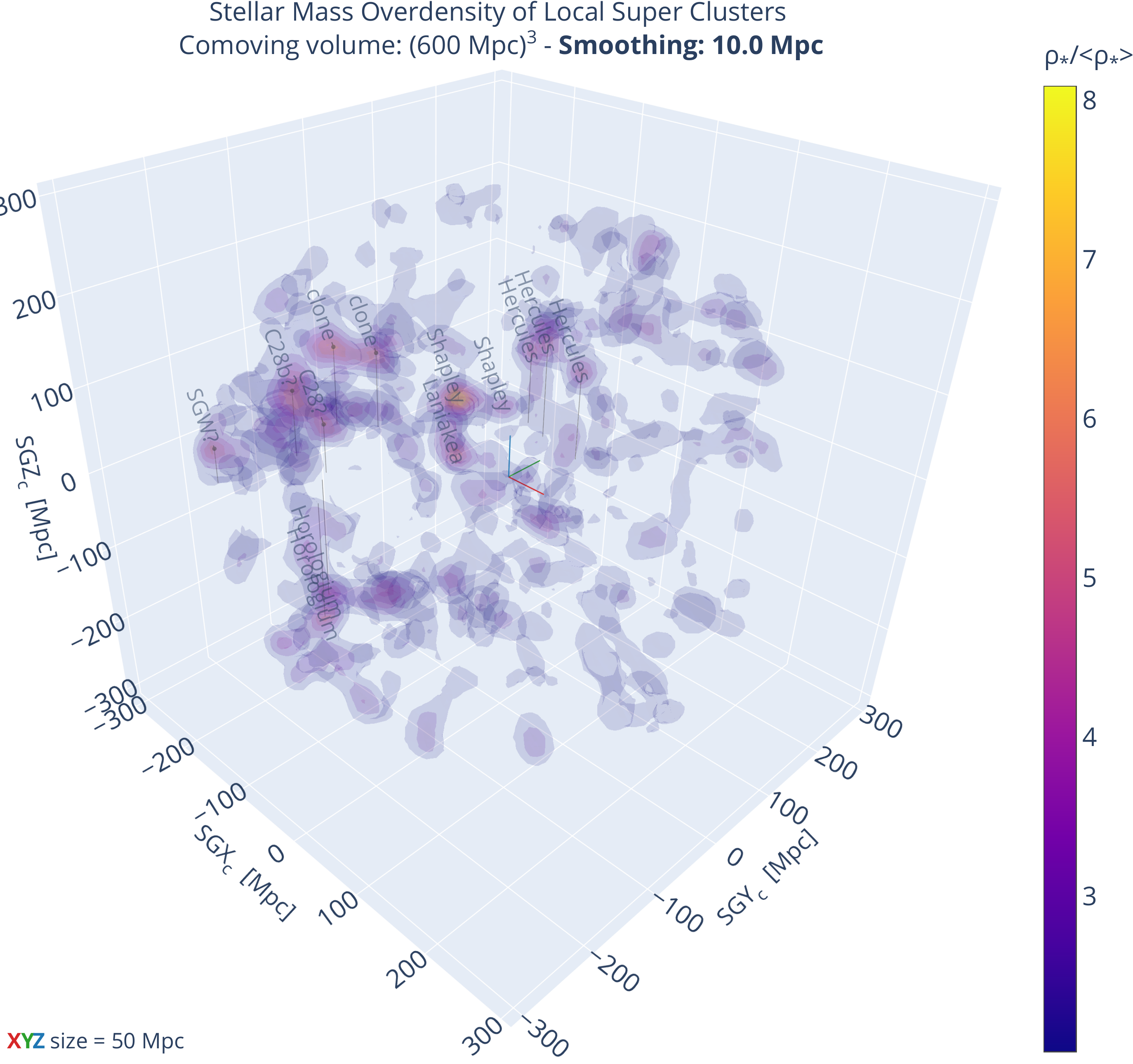}
    \caption{The stellar-mass overdensity in the Local Sheet (\textit{left}) and in the Local Universe (\textit{right}). The density fields displayed in comoving supergalactic coordinates are smoothed with a 3D Gaussian filter on comoving scales of 0.2\,Mpc (\textit{left}) and 10\,Mpc (\textit{right}). See Ref.~\cite{2021arXiv210511345B} for more details.\vspace{-0.5cm}}
    \label{fig:3D}
\end{figure}

Figure~\ref{fig:3D} displays the 3D $M_*$ distribution inferred from the catalog, normalized to the mean value within 350\,Mpc. The distribution of galaxies within a box of 10\,Mpc-width in comoving supergalactic coordinates is consistent with a planar distribution along the Local Sheet or the supergalactic plane \cite{2014MNRAS.440..405M}. The $M_*$ distribution within a box of 600\,Mpc-width matches the 3D density fields inferred from Cosmic Flows (e.g.\ \cite{2017NatAs...1E..36H}). The Cosmic-Flow distribution of matter was shown to provide a satisfactory model to the arrival directions of UHECRs above 8\,EeV \cite{2021ApJ...913L..13D}, where a dipolar pattern is detected beyond $5\,\sigma$ confidence level \cite{2018ApJ...868....4A}. The authors of Ref.~\cite{2021ApJ...913L..13D} account in particular for the limited horizon of UHECRs induced by their propagation in the cosmic microwave and infrared backgrounds, as well as for magnetic deflections. Our catalog offers the opportunity not only to verify their results on large angular scales at energies larger than 8\,EeV but also, as per the discrete nature of our catalog, to model the UHECR sky on smaller angular scales up to the highest energies, where the UHECR horizon shrinks. The present contribution focuses on accounting for the UHECR horizon and on investigating $B$-fields on cluster scales.\vspace{-0.1cm}

\section{The limited reach of UHECRs}\vspace{-0.2cm}

We model the flux and composition of UHECRs reaching Earth following an approach similar to that developed in Ref.~\cite{2017JCAP...04..038A} (``SPG'' reference model). The three essential ingredients are the 1D distribution of the UHECR production rate, the UHECR spectrum at escape from the source environment and the modeling of energy losses and photo-dissociation of UHECRs along the line of sight. For the latter ingredient, we exploit \texttt{Simprop} \cite{2017JCAP...11..009A} to generate a transfer tensor tabulating the number of UHECRs reaching Earth as a function of observed energy and nuclear species, for the initial injection of a given nuclear species at redshift $z$ and initial rigidity $R$ (energy over charge). The tensor is stored as a \texttt{numpy} 5D array, whose product with any source distribution as a function of $z$ and injection spectrum as a function of $R$ can be efficiently computed. The injection spectrum follows a power-law with a rigidity cutoff \cite{2017JCAP...11..009A} and is assumed to be universal, that is independent of the host galaxy and identical for each injected nuclear species. The latter hypothesis follows the so-called Peters' cycles, i.e.\ acceleration up to a maximum energy proportional to the charge.

\begin{figure}[t]
    \centering
    \includegraphics[width=.49\textwidth]{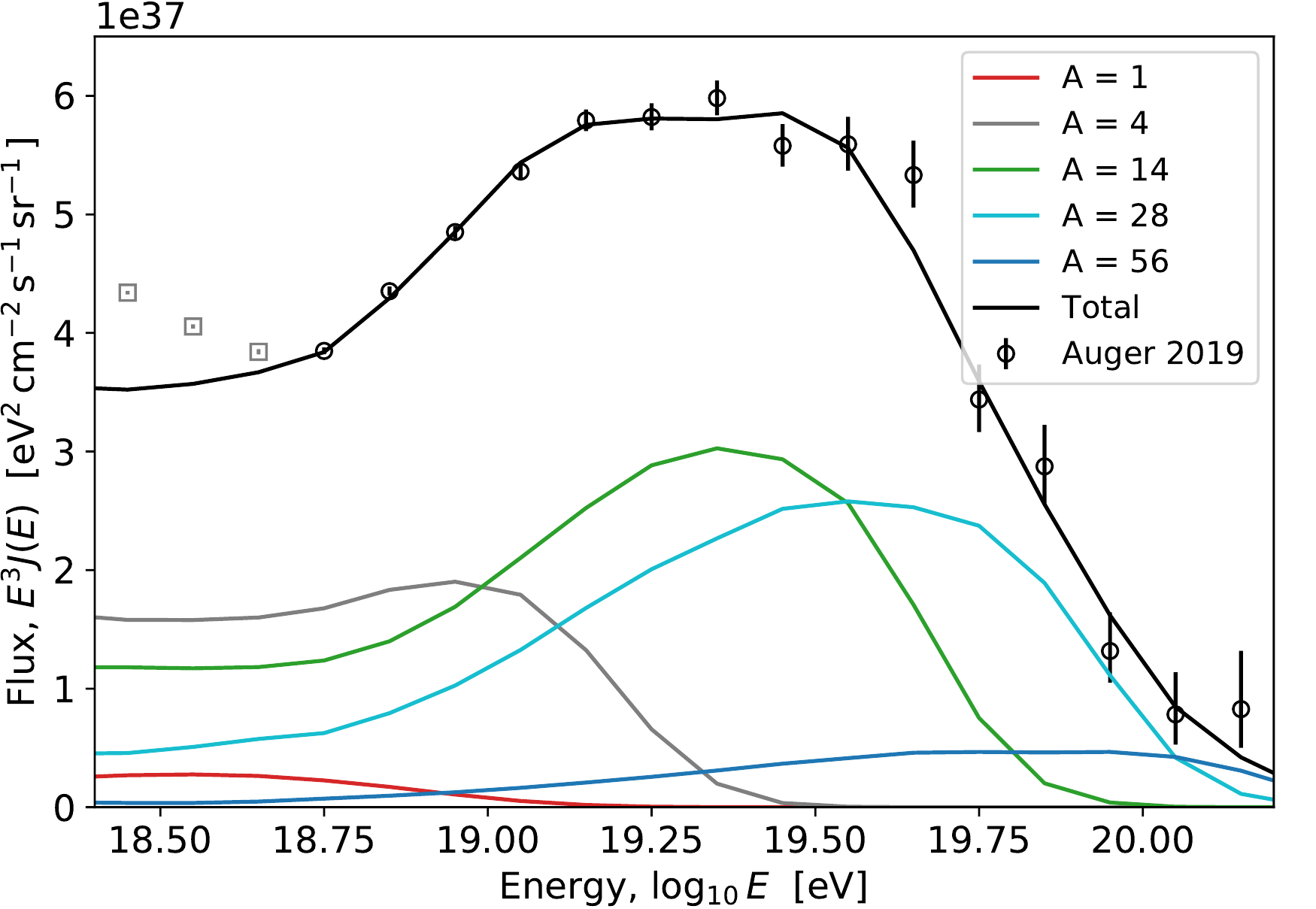} \hfill \includegraphics[width=.49\textwidth]{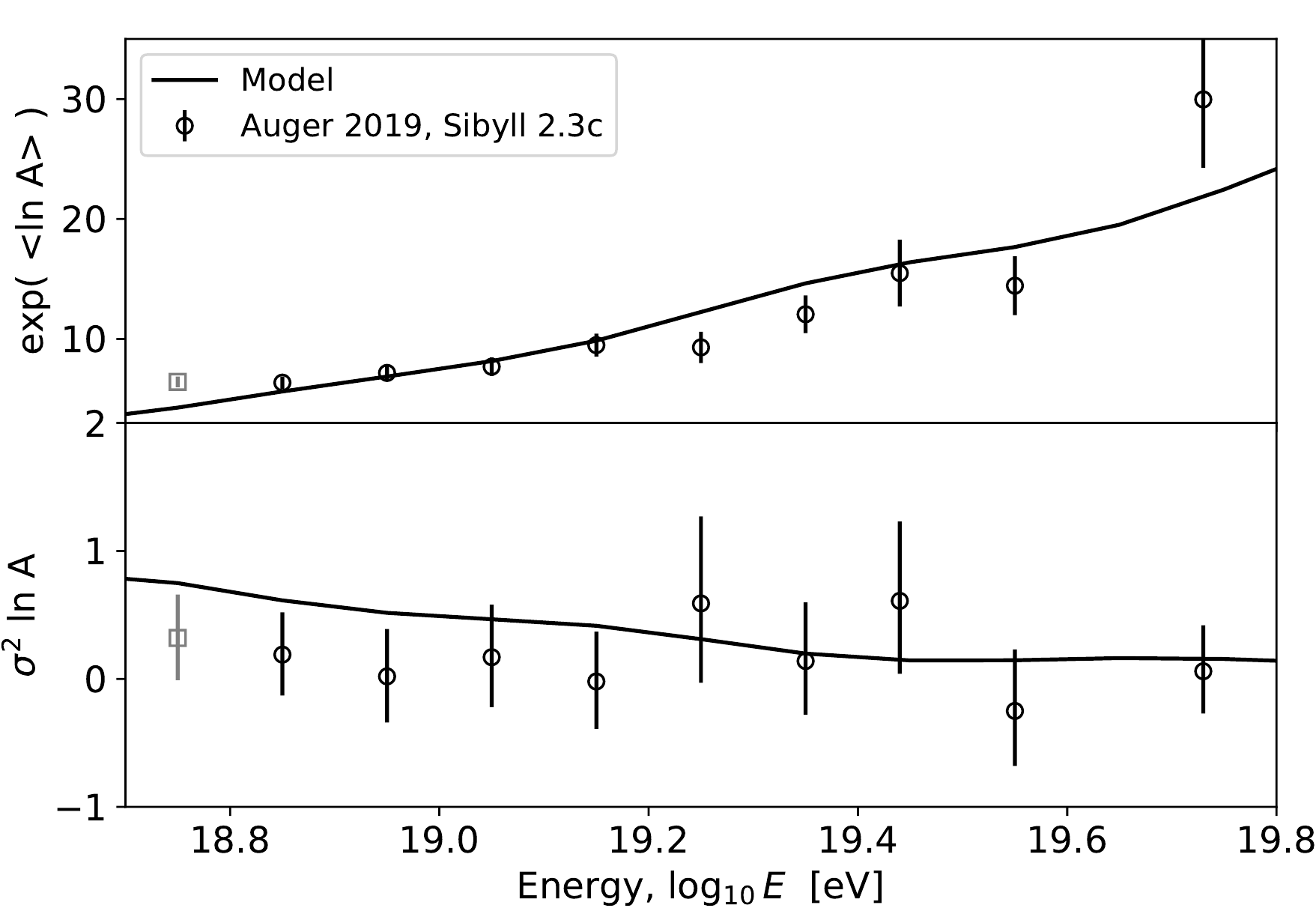}
    \caption{The combined fit of the UHECR spectrum and mass composition observables. The flux contributions from five injected nuclear species (H, He, N, Si, Fe), illustrated by colored lines, are fit to the spectral data from Ref.~\cite{2020PhRvL.125l1106A} and to the 1st and 2nd moments of the mass estimator, $\ln A$, from Ref.~\cite{Yushkov:2019J8}. The points shown as gray squares are not included in the fit and black circles display the observables accounted for.\vspace{-0.5cm}}
    \label{fig:UHECRobs}
\end{figure}

Indications of anisotropies at the highest energies \cite{2018ApJ...853L..29A} suggest that the UHECR production rate could be traced by SFR, in line with a scenario where UHECRs are accelerated in the late stages of the life of young massive stars, such as long GRBs. Such transient events could appear as persistent at ultra-high energies because of the temporal spread induced by intervening $B$-fields. We adopt the SFR density illustrated in Fig.~\ref{fig:sfrd} to account for the local distribution of matter beyond the Local Group. We developed a \texttt{python} software to jointly model the public spectrum \cite{2020PhRvL.125l1106A} and mass composition (1st and 2nd order moments of $\ln A$, \cite{Yushkov:2019J8}) reconstructed by the Pierre Auger Collaboration. While the UHECR spectrum is now reconstructed with exquisite precision from the Pierre Auger Observatory up to declinations $\delta < 25^\circ$, mass composition depends on the adopted hadronic interaction model, resulting in an overall systematic uncertainty on the mass scale. We adopt the mass observables inferred from Sybill 2.3c, for which a good fit of the data is found. Figure~\ref{fig:UHECRobs} illustrates the best-fit model obtained by injecting five nuclear species with a hard escape index of $-1.4$ and a rigidity cutoff at $R_{\rm max} = 10^{18.2}\,$V from sources following the SFR density at $\delta < 25^\circ$. Despite differences in the source distribution, the parameters of the escape spectrum are comparable to the index of $-1.5$ and cutoff at $10^{18.3}\,$V found by the Pierre Auger Collaboration (\cite{2017JCAP...04..038A}, Sybill 2.1). Nonetheless, the SFR distribution of galaxies in our catalog challenges the assumption of a smooth or nearly constant evolution within 1\,Gyr, as sometimes adopted in the literature. The inhomogeneity of the Local Universe could result in different spectra for different sky coverages at the highest energies, which could help to interpret differences between observations from the Telescope Array and Pierre Auger Observatory in different declination bands. A more quantitative assessment of this effect will be developed in an upcoming publication.

\begin{figure}[t]
    \centering
    \includegraphics[width=.75
    \textwidth]{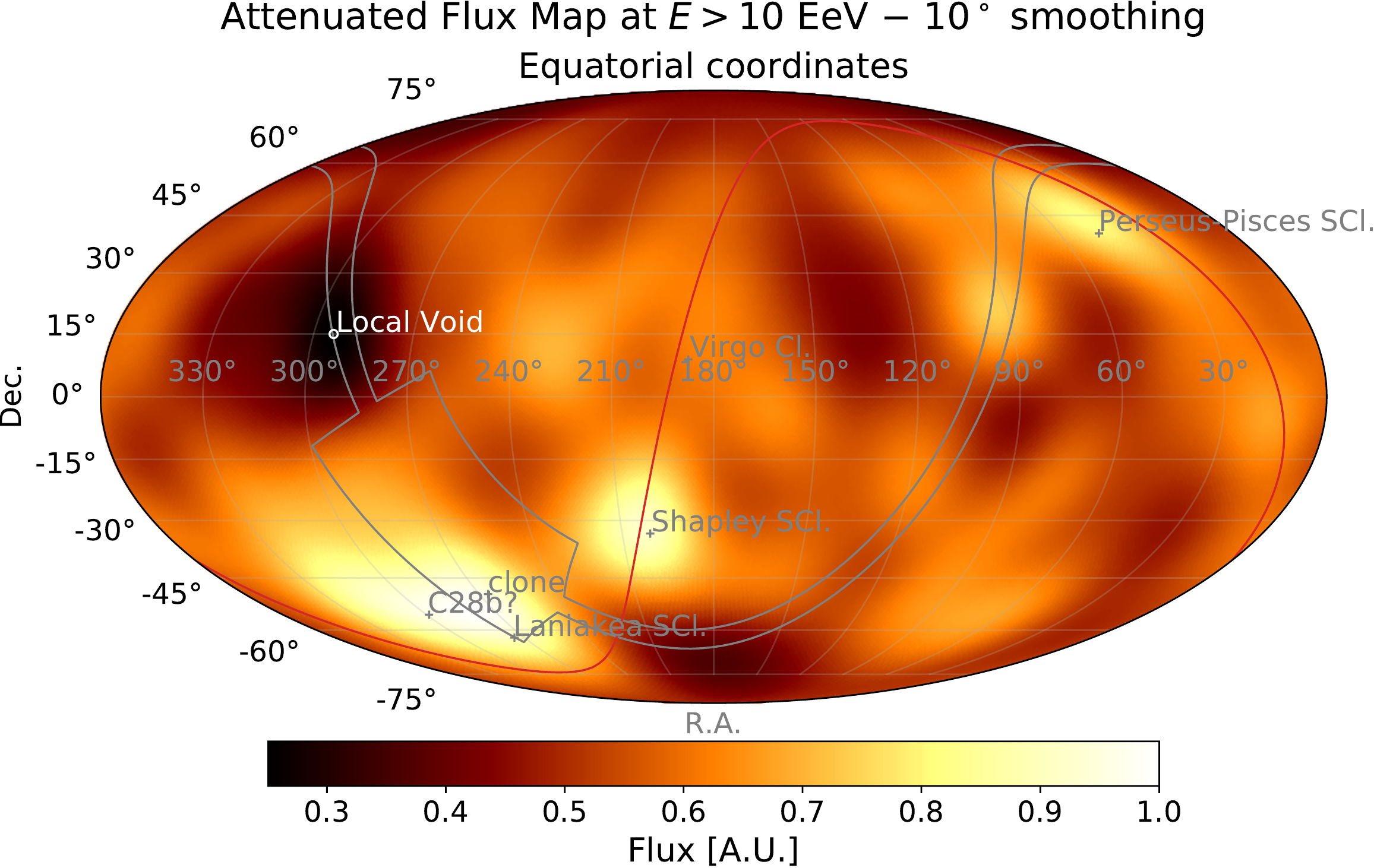}\\ \vspace{0.15cm} \includegraphics[width=.75\textwidth]{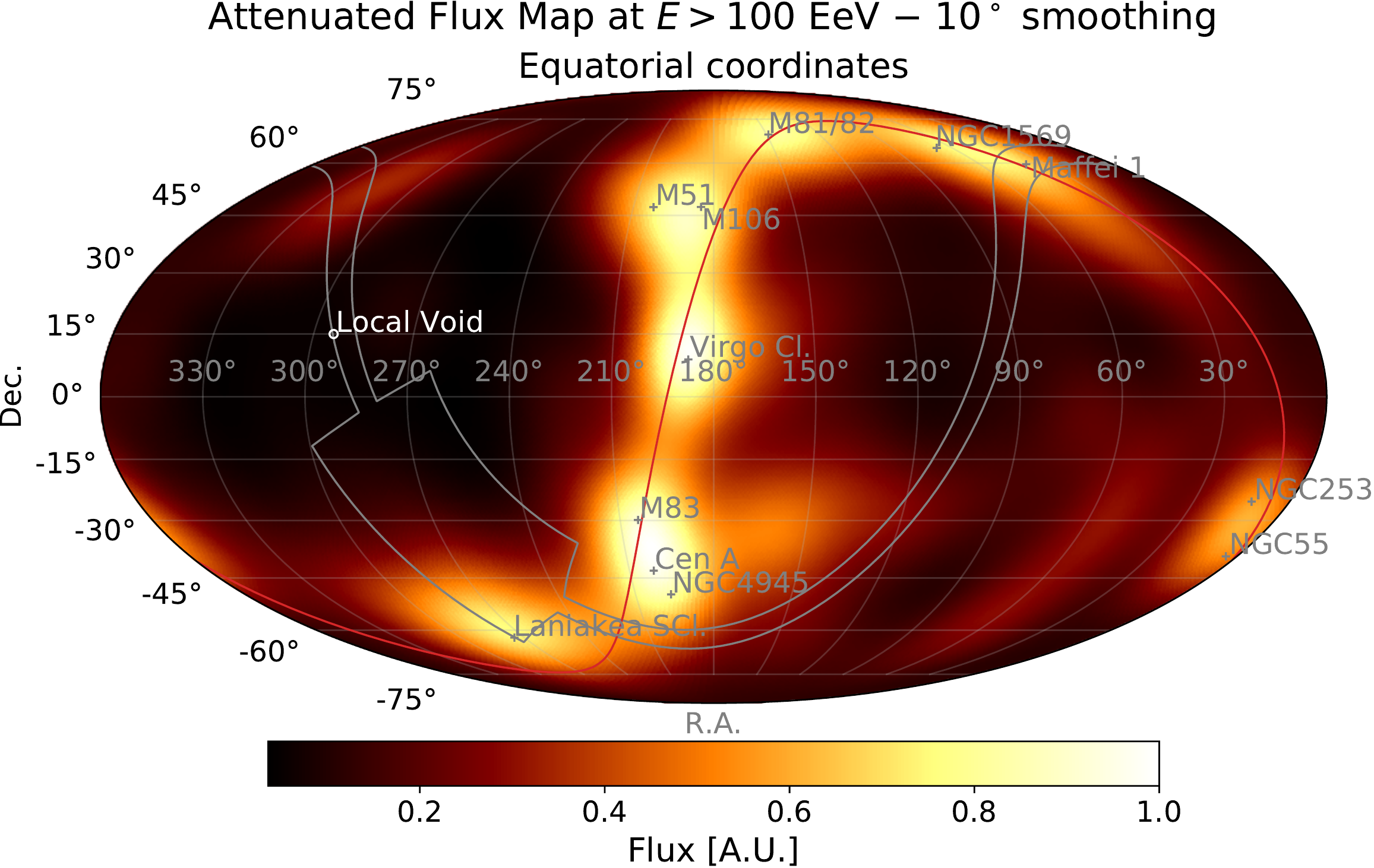}
    \caption{The sky maps, in equatorial coordinates, of the flux expected from star forming galaxies beyond the Local Group, their host clusters (Cl.)\ and superclusters (SCl.) accounting for UHECR attenuation above 10 EeV (\textit{top}) and 100 EeV (\textit{bottom}). The maps are smoothed on a 10$^\circ$ angular scale. The supergalactic plane and the Zone of Avoidance are displayed as a red line and gray contour, respectively.\vspace{-0.5cm}}
    \label{fig:skymaps}
\end{figure}

The skymaps expected at two extreme ends of the UHECR spectrum above the ankle are displayed in Fig.~\ref{fig:skymaps}. Both maps are smoothed on a Gaussian angular scale of $10^\circ$ to enable a qualitative comparison with UHECR observations (e.g.\ \cite{2019EPJWC.21001005B}). 
It should be noted that the energy thresholds adopted in Fig.~\ref{fig:skymaps} differ from those in Ref.~\cite{2019EPJWC.21001005B}. As UHECR propagation strongly depends on rigidity, UHECR maps above a fixed energy threshold are expected to be strongly dependent on the maximum rigidity cutoff at escape from the sources, which in turn can vary by a factor of about two depending on the selected hadronic interaction model \cite{2017JCAP...04..038A}. Adopting the mass-composition scale as a free nuisance parameter  will help to determine the ability of our model to reproduce the observed UHECR sky as a function of energy.

Interesting similarities between the model and UHECR observations can be noted. The model of the UHECR sky above 10\,EeV is dominated by structures on hundred-Mpc scales, resulting in a more isotropic flux distribution than that inferred at higher energies. The UHECR model beyond 100\,EeV displays overdensities around the supergalactic plane, as expected from the very galaxies in the Local Sheet that are inferred to be responsible for the indication of anisotropies at the $4\,\sigma$ confidence level at energies $\gtrsim 40\,$EeV \cite{2018ApJ...853L..29A}. The main underdensity in both maps is due to the Local Void, whose direction coincides well the UHECR anti-dipole direction above 8\,EeV. The model also matches the other underdensity at the highest energy, west of the supergalactic plane.

As is often the case, discrepancies between model and observations may carry more substantial information than similarities do. The largest UHECR overdensities above 10\,EeV are expected from major superclusters: our own, Laniakea, and its companion, Perseus-Pisces, whose peak densities are ${\sim}\,70\,$Mpc away from us, as well as the Shapley and C28 concentrations at distances of ${\sim}\,200-250\,$Mpc. The close-by outskirts of Laniakea as well as the Virgo cluster ($\delta \approx 0^\circ$), at a distance of ${\sim}\,16\,$Mpc, are expected to contribute to the UHECR sky at the highest energies. None of these overdensities are apparent in the observed UHECR skymaps \cite{2019EPJWC.21001005B}.\vspace{-0.1cm}

\section{The impact of magnetic fields on cluster scales}\label{sec:cluser_fields}\vspace{-0.2cm}

\begin{figure}[b]
    \centering
    \includegraphics[height=5.3cm]{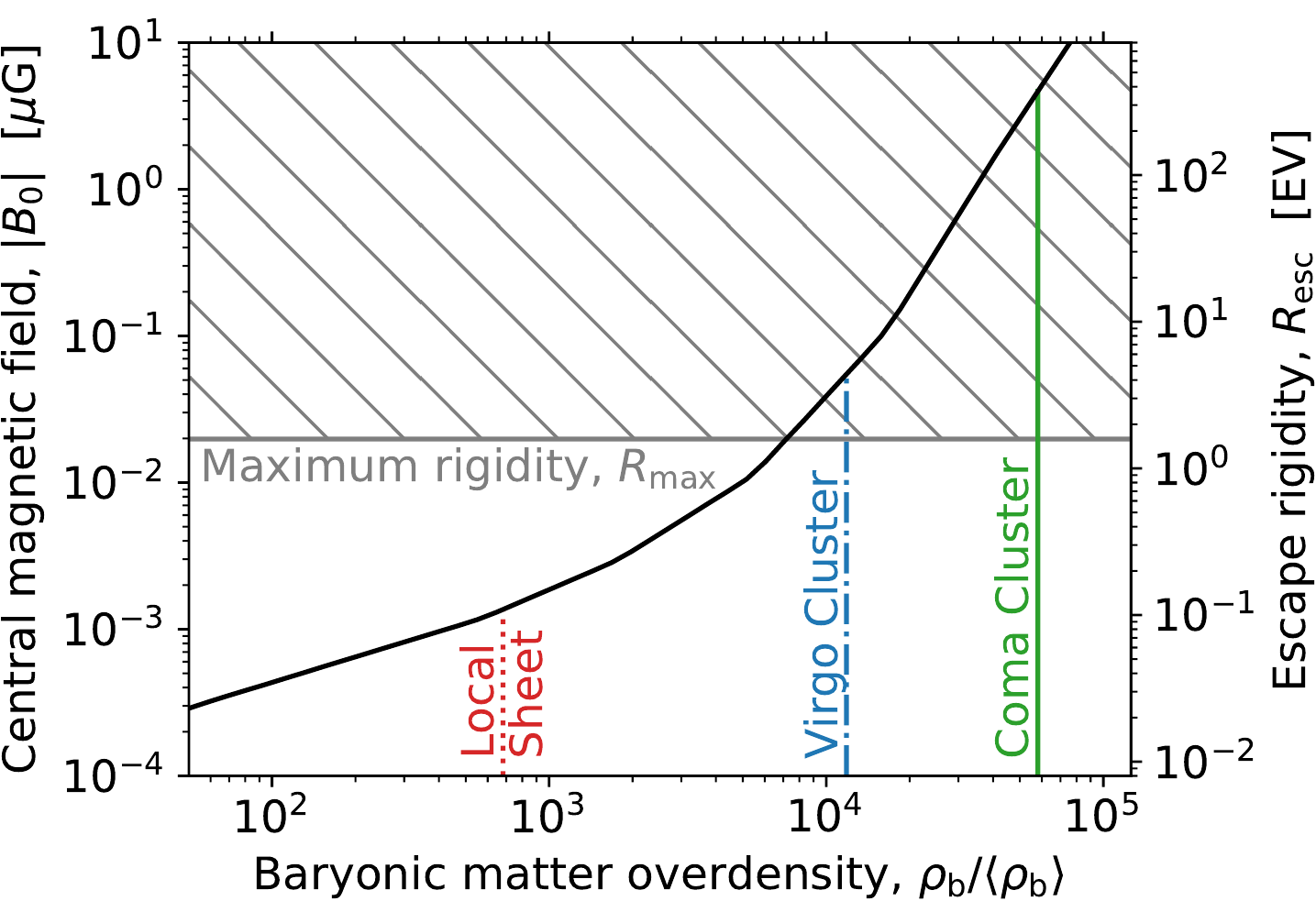} \hfill \includegraphics[height=5.3cm]{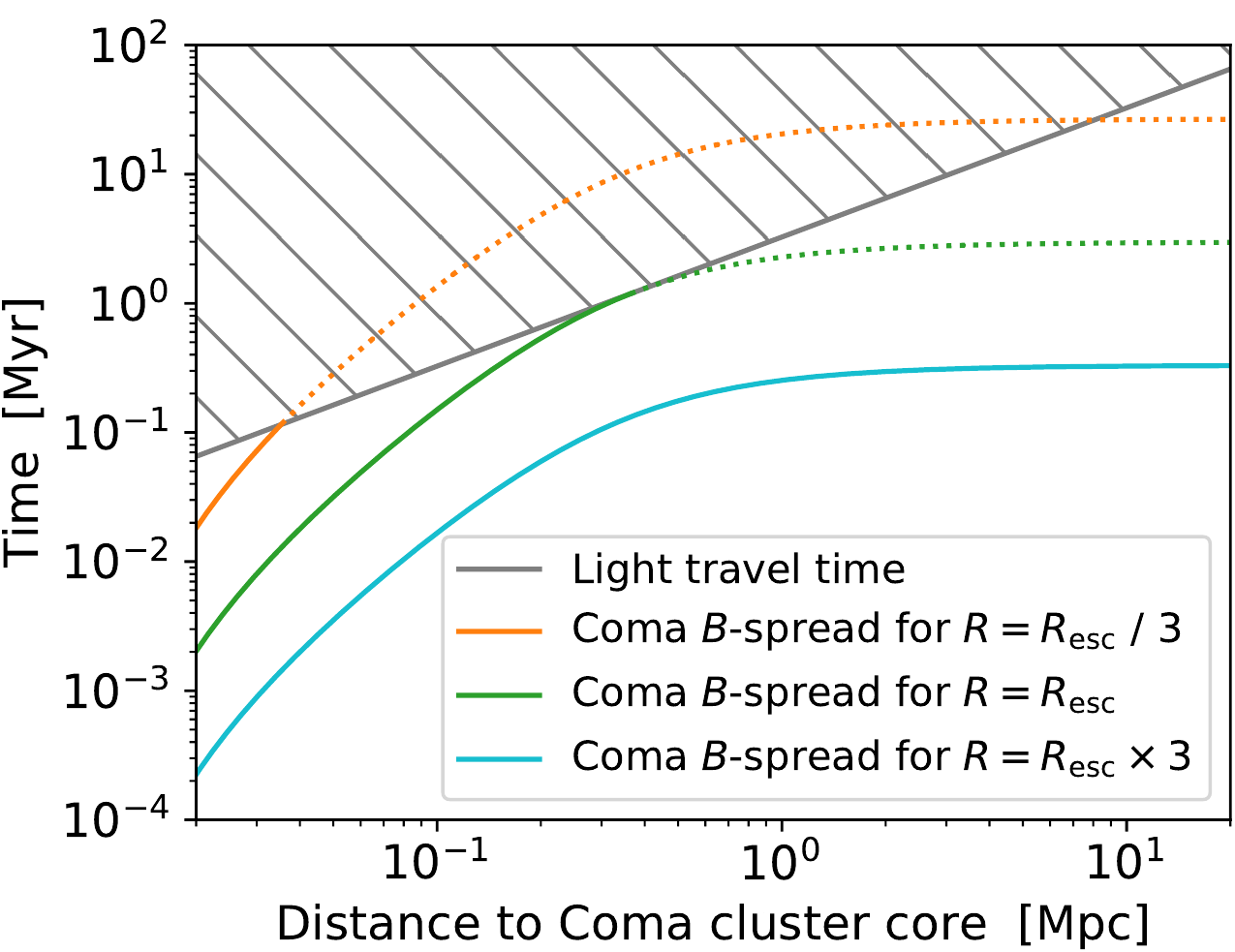}
    \caption{\textit{Left:} The central magnetic field on cluster scales as a function of baryonic overdensity. The model in black is extracted from Ref.~\cite{2018SSRv..214..122D}. The maximum rigidity inferred from UHECR data is illustrated by a hashed region. \textit{Right:} UHECR time spread induced by propagation away from the core of the Coma cluster. The light travel time is illustrated by a hashed region. Dotted lines figure UHECR confinement.\vspace{-0.5cm}}
    \label{fig:escape}
\end{figure}

One way out of the conundrum would consist in invoking Galactic $B$-fields (see e.g.\ \cite{2019ICRC...36..436T} for an overview of their effects). A specific configuration of its regular component could result in shadowing the unobserved overdensities while keeping intact the favorable properties of the model. We do not explore this road and rather focus on $B$-fields expected on cluster scales. Magneto-hydrodynamics cosmological simulations predict a scaling of magnetic-field amplitude with baryonic overdensity, illustrated in Fig.~\ref{fig:escape} for the only model from Ref.~\cite{2018SSRv..214..122D} able to reach $\mu$G amplitude at cluster centers. Such amplitudes are observed e.g.\ in the Coma cluster (central value of $4.7 \pm 0.7\,\mu$G), whose $B$-field properties are constrained in Ref.~\cite{2010A&A...513A..30B}. 

We estimate baryonic overdensities, $\rho_{\rm b}/\langle \rho_{\rm b} \rangle$, adopting the scaling of baryonic mass with $M_*$ from Ref.~\cite{2013ApJ...778...14G}. As in Fig.~\ref{fig:3D}, The 3D volume of 350\,Mpc is smoothed on a Gaussian scale of 200\,kpc, roughly matching that of clusters. We estimate $\rho_{\rm b}/\langle \rho_{\rm b} \rangle \approx 20,000$ for Coma, which would correspond to a central $B$-field of $0.14\,\mu$G following Ref.~\cite{2018SSRv..214..122D}. Such a mismatch in $B$-field by a factor ${\sim}\,30$ with respect to observations can be mitigated by rescaling the baryonic overdensity by a factor of ${\sim}\,3$ as illustrated in Fig.~\ref{fig:escape}, left. The impact of the Coma $B$-field on UHECR propagation can be estimated from its average coherence length and spatial extent provided in Ref.~\cite{2013ApJ...778...14G}. We estimate the minimum rigidity, $R_{\rm esc}$, that UHECRs should reach to escape Coma as the magnetic horizon of the structure, i.e.\ by equating the light travel time from its center to the time delay induced in the regime of weak UHECR deflections \cite{2008PhRvD..78b3005M}, as illustrated in Fig.~\ref{fig:escape}, right. The escape rigidity is a linear function of the central $B$-field, as shown on right-hand-side y-axis of Fig.~\ref{fig:escape}, left.

The escape rigidity exceeds the maximum one inferred from the combined fit, so that UHECR emitted in such dense clusters should remain confined and could never reach Earth. The overdensities estimated for the Virgo cluster and the Local Sheet yield central $B$-fields of ${\sim}\,50\,$nG  and ${\sim}\,1\,$nG, respectively, in line with those inferred from simulations \cite{2018MNRAS.475.2519H}. Under the approximation of a coherence length and extent similar to those of Coma, the $B$-field of the Virgo cluster should result in a blind spot in the UHECR sky, possibly reconciling our model in Fig.~\ref{fig:skymaps} with observations. A suppression at the the low end of the UHECR escape spectrum is expected for numerous galaxies in our catalog. Interestingly, the same argument applied to the Local Sheet suggests that UHECRs of rigidities smaller than ${\sim}\,0.1\,$EV cannot enter our halo. $B$-fields on cluster scales could thus provide an explanation for the apparent hard escape index. More advanced studies (see e.g. \cite{2013JCAP...10..013M}) will be developed to confirm the impact of $B$-fields on the spectrum at escape from host clusters.

Although $B$-fields on cluster scales remain underconstrained, they appear to be a crucial ingredient for the emerging UHECR astronomy. Our understanding of these $B$-fields could significantly benefit from X-ray and radio observations with eROSITA and SKA. Finally, a satisfactory UHECR model should explore the flux of secondary neutral particles induced by confinement on cluster scales \cite{2018NatPh..14..396F}. The possibility of detecting UHECR relics with next-generation neutrino and $\gamma$-ray instruments, such as KM3NeT/IceCube-Gen2 and CTA, should certainly be explored.  \vspace{-0.24cm}


\end{document}